\documentclass{acm_proc_article-sp}

\newcommand{\ignore}[1]{}

\newcommand{\boxtheorem}{\hfill $\Box$}
\newcommand{\nit}[1]{{\it #1}}

\usepackage{pdfsync}
\usepackage{verbatim}
\usepackage{hhline}

\usepackage{color}

\ignore{  }

\newcounter{lemma-counter}
\newcounter{example-counter}
\newcounter{proposition-counter}
{\vskip \abovedisplayskip \refstepcounter{lemma-counter}%
\noindent {\bf Lemma \arabic{lemma-counter}.}}%

\newenvironment{proposition}%
{\vskip \abovedisplayskip \refstepcounter{proposition-counter}%
\noindent {\bf Proposition \arabic{proposition-counter}.}}%

\newenvironment{example}%
{\vskip \abovedisplayskip \refstepcounter{example-counter}%
\noindent {\bf Example \arabic{example-counter}.}}%

\def\verbatim@font{\rmfamily\small}
\makeatother

\newcommand{\mc}[1]{\mathcal{ #1}}

\title{\vspace*{-5mm} Unifying Causality, Diagnosis, Repairs and View-Updates in Databases}

\numberofauthors{2} 
\author{
\alignauthor
Leopoldo Bertossi \\
      \affaddr{Carleton University}\\
       \affaddr{School of Computer Science}\\
       \affaddr{Ottawa, Canada}\\
       \email{\small bertossi@scs.carleton.ca}\vspace*{-1cm}
\alignauthor
Babak Salimi \\   \affaddr{Carleton University}\\
       \affaddr{School of Computer Science}\\
       \affaddr{Ottawa, Canada}\\
       \email{\small bsalimi@scs.carleton.ca}
}

\begin{document}
\bibliographystyle{plain}

\maketitle
\vspace{-1.5mm}

\begin{abstract}
In this work we establish and point out connections between the notion of query-answer causality in databases and
database repairs, model-based diagnosis in its consistency-based and abductive versions, and database updates through views.
The mutual relationships among these areas of data management and knowledge representation shed light on each of them and
help to share notions and results they have in common. In one way or another, these are all approaches to uncertainty management, which becomes
even more relevant in the context of big data that have to be made sense of.
\end{abstract}

\section{Introduction}

\vspace{1mm}
 Causality is not only a deep subject that appears at the foundations of many scientific disciplines, but also something we want to represent and compute in order to deal with the {\em uncertainty} of data, information and theories.  In data management, the need to understand and compute
why certain (query) results are obtained or not, or why certain natural semantic conditions are not satisfied can only grow and become more complex  when confronted with big data.
{\em It is difficult to make sense of the uncertainty associated to big data}, and causality is a fundamental and systematic way to confront the problem.

Our current  research is motivated by  trying to understand {\em causality in data management} from different perspectives. As described below, there are fruitful connections among fours forms of reasoning: inferring causes from  databases, model-based diagnosis, consistent query answering (and repairs), and view updates. They all reflect some sort of {\em uncertainty} about the information at hand.

When querying a database, a user may not obtain the expected results, and the system could provide some explanations. They could be useful to further understand the data or reconsider the
query. A notion of causality-based explanation for a query result was introduced  in \cite{Meliou2010a}.

Intuitively, a tuple $t$  is an {\em actual cause} for an answer $\bar{a}$ to a
conjunctive query $\mc{Q}$ from  a relational database instance $D$ if there is a ``contingent" set of tuples $\Gamma$,
such that, after removing $\Gamma$ from $D$, removing/inserting $t$ from/into $D$ causes $\bar{a}$ to switch from being an answer to being a non-answer.
Actual causes and contingent tuples  are restricted to be among a pre-specified set
of {\em endogenous tuples}, which are admissible, possible candidates for causes, as opposed to {\em exogenous tuples}. \ (For non-causality-based explanations for query answers in DL ontologies, see \cite{Borgida08}.)

Since some causes may be stronger than others, \cite{Meliou2010a} also introduces and investigates {\em responsibility},  that reflects the relative degree of actual causality. In applications involving large
data sets, it is crucial to rank potential causes according to their responsibilities \cite{Meliou2010b,Meliou2010a}.

Actual causation and responsibility, as used in  \cite{Meliou2010a}, can be traced back  to \cite{Halpern05,cs-AI-0312038}.\ignore{\footnote{$A$ is a cause of $B$ if, had $A$ not happened (this is the counterfactual
condition, since $A$ did in fact happen) then $B$ would not have happened}}
\ignore{Apart from the explicit use of causality, research on explanations for query results has focused mainly, and rather implicitly, on provenance
\cite{BunemanT07,Cheney09,tannen}, and
more recently, on provenance for non-answers \cite{ChapmanJ09,HuangCDN08}. } For connections between
causality and provenance, see \cite{Meliou2010a,Meliou2010b}. \ignore{However, causality is a more refined notion that identifies causes
for query results on the basis of  user-defined criteria, and ranks causes according to
their responsibility \cite{Meliou2010b}. }

 Model-based diagnosis, and {\em consistency-based diagnosis} in particular \cite{Reiter87}, is an area of knowledge representation. A {\em system specification} in a logical formalism and an a system's {\em observation} are given. Typically, the specification tells us how the system works under normal conditions, the observation
 is unexpected under those conditions, and an {\em explanation} for the {\em failure} is required.

In a different direction, a database instance $D$ may not satisfy certain intended integrity constraints (ICs). A {\em repair} of $D$ is a
database   $D'$ that does satisfy the ICs and {\em minimally departs} from $D$. Different forms of minimality have been investigated. A {\em consistent answer} to a query
$\mc{Q}$ from $D$ wrt. the ICs is an answer to $\mc{Q}$ that is obtained from all possible repairs, i.e. is {\em invariant} or {\em certain} under the class of repairs (see \cite{2011Bertossi} for a recent survey). (Not in the framework of repairs, consistency-based diagnosis techniques have been applied
to restoring a DB from IC violations \cite{Gertz97}.)

Interestingly, deeper and useful connections between these areas
(and others, see below) are starting to emerge. Actually,  we have reported in \cite{nmr14}, where more results and details can be found,   on new results about precise connections
between  causality for query answers in databases, database repairs wrt. denial constraints, and consistency-based diagnosis.

More precisely, it is possible to obtain database repairs from causes, and the other way around. Then, the vast
body of research on database repairs can be applied to the newer problem of determining actual causes for query answers.
We unveil a
strong connection between computing causes and their responsibilities for conjunctive queries, on the one hand, and computing both {\em subset-based} and {\em cardinality-based repairs} in databases
\cite{icdt07}
wrt. denial constraints, on the other hand. These computational
problems can be reduced to each other.
Some results about this connection are briefly presented
in Section \ref{sec:causes}.

Furthermore, by formulating the causality problem as a diagnosis problem, it is possible to characterize causes in terms of the system's diagnoses. Actually, inferring and computing actual causes  and responsibility in a database setting become, in different forms,
consistency-based diagnosis reasoning problems and tasks.
More specifically,
a causal explanation for a conjunctive query answer can be viewed as a diagnosis, where in essence  the relational database
 provides the system description, and the observation is the query answer. Some results are summarized in Section
 \ref{sec:diagnosis}.

Abduction, as another approach to model-based diagnosis, can also be used to characterize causes for answers to queries, in particular
Datalog queries. We present some results on this connection in Section \ref{sec:abduction}.

We conclude this overview of mutual relationships between database causality and other areas of data management and knowledge representation by
making some very general remarks, in Section \ref{sec:views}, on database updates through views. We point out several  connections with the above mentioned
areas and problems.

\section{Causes and Repairs}\label{sec:causes}
\vspace{1mm}

 We assume that a relational instance is split in two subinstances, say $D=D^n \cup D^x$,  where $D^n$ and $D^x$ are formed by {\em  endogenous} and
 {\em exogenous} tuples,
 respectively. The former tuples are  possible candidate for causes. Now, given a boolean conjunctive $\mc{Q}$, a tuple $t \in D^n$ is a
{\em counterfactual cause} for $\mc{Q}$  if $D\models \mc{Q}$ and $D\smallsetminus \{t\}  \not \models \mc{Q}$.
A tuple $t \in D^n$ is an {\em actual cause} for  $\mc{Q}$,
if there  exists $\Gamma \subseteq D^n$, called a {\em contingency set},  such that $t$ is a counterfactual cause for $\mc{Q}$ in $D\smallsetminus \Gamma$
\cite{Meliou2010b}.

   The numerical function, {\em responsibility}, reflects the relative degree of causality of a tuple for
a query result. More precisely, the {\em responsibility} of an actual cause $t$ for $\mc{Q}$, denoted by $\rho(t)$,  is the numerical value $\frac{1}{(|\Gamma| + 1)}$, where $|\Gamma|$ is the
size of the smallest contingency set for $t$ (with minimum cardinality). Tuples with higher responsibility are considered  to provide more interesting explanations for query results \cite{Meliou2010b,Meliou2010a}.

We will denote with $\mc{CS}(D^n,D^x,\mc{Q})$ the set of actual causes for $\mc{Q}$ (being true) in instance $D=D^n \cup D^x$. The following (anti)monotonicity results immediately
hold.

\vspace{-3mm}
\begin{proposition}
Let $(D^n)^\prime, (D^x)^\prime$ denote updates of $D^n,D^x$ by insertion of tuple $t$, resp.
\ It holds: \ (a)  $\mc{CS}(D^n,D^x,$ $\mc{Q}) \ \subseteq \ \mc{CS}((D^n)^\prime,D^x,\mc{Q})$. \  (b)  $\mc{CS}(D^n, (D^x)^\prime, $ $\mc{Q}) \ \subseteq \ \mc{CS}(D^n,D^x,\mc{Q})$. \boxtheorem
\end{proposition}
Actually, the set of actual causes may shrink by adding an exogenous tuple \cite{nmr14}.
The possible loss of actual causes, which shows a non-monotonic behavior, is in line with the connections of  causality with database repairs and model-based diagnosis, both of which have
associated reasoning tasks that are also non-monotonic.

Now assume that the query is of the form \ $\mc{Q}\!: \exists \bar{x}(P_1(\bar{x}_1) \wedge \cdots \wedge P_m(\bar{x}_m))$, with $\bar{x} = \cup \bar{x}_i$, and
$\mc{Q}$ is unexpectedly true in  $D$, i.e we expected $D \models \neg \mc{Q}$. Then, we may want to trace back the causes for $\mc{Q}$ to be true.

We first notice that $\neg \mc{Q}$ is logically equivalent to  the {\em denial constraint} \ (DC) \ $\kappa(\mc{Q})\!: \forall \bar{x} \neg (P_1(\bar{x}_1) \wedge \cdots \wedge P_m(\bar{x}_m))$ (that is also sometimes written as the Datalog constraint, \ $\leftarrow P_1(\bar{x}_1),\ldots,P_m(\bar{x}_m)$).
If $\mc{Q}$ is not expected to hold, we may consider $D$ to be inconsistent wrt. $\kappa(\mc{Q})$. Repairs of $D$ wrt. $\kappa(\mc{Q})$ may be considered.

More precisely,  we first consider the {\em S-repairs} (aka. {\em subset-repairs}), which are those consistent instances obtained from
$D$ via tuple insertion/deletion and make the symmetric difference with $D$  minimal wrt. set inclusion. In the case of DCs, S-repairs are subsets of $D$  that do not have any
proper subset that is a repair \cite{2011Bertossi}. Next, we consider the class containing the set differences between $D$ and those {\em S-repairs}  that do not contain tuple $t \in D^n$, and are obtained
by removing a subset of $D^n$:
\begin{eqnarray}
\mc{DF}(D, D^n,\kappa(\mc{Q}), t)\!\!\!\!&=&\!\!\!\!\{ D \smallsetminus D'~|~ D' \in \nit{Srep}(D,\kappa(\mc{Q})), \nonumber\\
&&~~~~~~~~~~t \in (D\smallsetminus D') \subseteq D^n\}. \label{eq:reps}
\end{eqnarray}
Here, $\nit{Srep}(D,\kappa(\mc{Q}))$ denotes the class of S-repairs of instance $D$ wrt.  $\kappa(\mc{Q})$.

\vspace{-3mm}
\begin{proposition}  Given $D= D^n \cup D^x$, a BCQ  $\mc{Q}$, and $t \in D^n$:\\
(a) $t$ is an actual cause for $\mc{Q}$ iff
$\mc{DF}(D, D^n,\kappa(\mc{Q}), t)$ $\neq \emptyset$.\\
(b) If $\mc{DF}(D, D^n, \kappa(\mc{Q}),  t) = \emptyset$, then $\rho(t)=0$.\\
(c) If $\rho(t)\neq 0$, $\rho(t)=\frac{1}{|s|}$, where $s \in \mc{DF}(D,D^n, \kappa(\mc{Q}), t)$
and there is no $s' \in \mc{DF}(D, D^n,\kappa(\mc{Q}), t)$ such that, $|s'| < |s|$.
\boxtheorem
\end{proposition}

In the other direction, it is also possible to obtain repairs from actual causes.
In fact, consider the database instance $D$  and the denial constraint $\kappa\!: \ \leftarrow A_1(\bar{x}_1),\ldots,A_n(\bar{x}_n)$.
As usual, a boolean conjunctive {\em  violation view}, $V^\kappa\!: \exists\bar{x}(A_1(\bar{x}_1)\wedge \cdots \wedge A_n(\bar{x}_n))$, can be associated to $\kappa$.

 Given an inconsistent instance $D$ wrt.  $\kappa$, we collect all S-minimal contingency sets associated with the actual cause $t$ for $V^\kappa$, as follows:

\noindent $\mc{CT}(D,D^n,V^\kappa,t) = \{  s\subseteq D^n~|~D\smallsetminus s \models V^\kappa,D\smallsetminus (s \cup \{t\}) \not \models$\\
\hspace*{2.7cm}$ V^\kappa,   \mbox{ and }
   \forall s''\subsetneqq s, \ D \smallsetminus (s'' \cup \{t\})  \models V^\kappa \}$.

\vspace{-4mm}
\begin{proposition}\label{pro:sr&cp}
(a)  $D$ is consistent wrt. $\kappa$ iff
$\mc{CS}(D, \emptyset, V^\kappa) = \emptyset$. \ (b)
$D' \subseteq D$ is an S-repair for $D$ iff,  for every $t \in D \smallsetminus D'$,
$t \in \mc{CS}(D, \emptyset, V^\kappa)$ and $D \smallsetminus (D' \cup \{t\}) \in \mc{CT}(D, D,V^\kappa, t)$. \boxtheorem
\end{proposition}
This proposition is stated for a single  DC. However, it is possible to obtain repairs from causes for larger sets of
DCs. In \cite{nmr14}
we provide a (naive) algorithm that computes the S-repairs for an inconsistent instance $D$ and  a set $\Sigma$ of DCs from causes for their violation views being
true.

Consider an instance $D$, and a DC $\kappa$. The following proposition
establishes a relationship between {\em consistent query answering} (CQA) wrt. the S-repair semantics \cite{2011Bertossi} and actual cases for the violation view $V^{\kappa}$.

\vspace{-3mm}
\begin{proposition} A ground atomic query $A$ is consistently true, denoted $D \models_S A$, iff $A \in D \smallsetminus \mc{CS}(D,\emptyset, V^{\kappa})$.\boxtheorem
\end{proposition}

Along a similar line,  {\em C-repairs}\footnote{These are {\em cardinality-repairs}, and are defined as the S-repairs, but they minimize the cardinality of the symmetric difference.
For DCs, they become maximal subsets, in cardinality, of the original instance \cite{2011Bertossi}.} are related to most responsible actual causes. We can collect the most responsible actual causes for $V^\kappa$:

\noindent $\mc{MRC}(D, V^\kappa)= \{t \in D~|~ t \in \mc{CS}(D,\emptyset,V^\kappa), \not \exists t' \in \mc{CS}(D,\emptyset,V^\kappa)$\\
\hspace*{3.5cm}$ \mbox{ with } \rho(t')> \rho(t)   \}$.

\vspace{-3mm}
\begin{proposition} For an instance $D$ and denial constraint $\kappa$, $D'$ is a C-repair for $D$ wrt. $\kappa$ iff  for each $t \in D \smallsetminus D'$:
$t \in  \mc{MRC}(D,V^\kappa)$ and $D \smallsetminus (D' \cup \{t\}) \in \mc{CT}(D, D,V^\kappa, t)$. \boxtheorem
\end{proposition}
In our ongoing research we have established and exploited a  close connection (to be reported somewhere else) between {\em consistent answers} -under the cardinality-based repair se-mantics- to conjunctive queries wrt. denial constraints and
most-responsible causes. These problems can be reduced to each other.

The partition of a database into endogenous and exogenous tuples has been exploited in the
context of causality. However, this kind of partition is also of interest in the context of repairs. Considering that we should
have more control on endogenous tuples than on exogenous ones, which may come from external sources, it makes sense to consider
{\em endogenous repairs} that are obtained by updates (of any kind) on endogenous tuples. For example, in the case of violation of denial constraints,
   endogenous repairs would be obtained -if possible- by deleting endogenous tuples only.

   If there are no repairs based on endogenous tuples
   only, a preference condition could be imposed on repairs, privileging those that change exogenous the least. (Of course,
   it could also be the other way around, i.e. we may feel more inclined to change exogenous tuples than our endogenous ones.)

   Actually, we could go even further and apply notions of {\em preferred repairs} \cite{chomicki12,ihab12}. If $\nit{Prep}$ denotes a given class
   of preferred repairs, it would be  possible to explore the use of a relationship as the one in (\ref{eq:reps}), replacing $\nit{Srep}$ by $\nit{Prep}$, and by doing so, define other notions of causes,
   say tuples that are $\nit{Pcauses}$ for query answers. Considering other, e.g. preference based, forms of causality was mentioned as an interesting open direction in \cite{Meliou2010a,Meliou2010b}.

   As a further extension, it could be possible to assume that combinations of (only) exogenous tuples never violate the ICs, something that could be checked
   at upload time. In this sense, there would be a part of the database that is considered to be consistent, while the other is subject to possible repairs.
   A situation like this has been considered, for other purposes and in a different form, in \cite{greco14}.

  Actually, going a bit further, we could even consider the relations  in the database with an extra, binary  attribute, $N$, that is used to annotate if a tuple is
  endogenous or exogenous (it could be both), e.g. a tuple like $R(a,b, \nit{yes})$. ICs could be annotated too, e.g. the ``exogenous" version of DC $\kappa$, could be
   $\kappa^E\!: \ \leftarrow P(x, y,\nit{yes}),R(y, z,\nit{yes})$, and could be assumed to be satisfied.

\section{Causes and Consistency-Based\\ Diagnosis}\label{sec:diagnosis}
\vspace{1mm}

As above, let $D = D^n\cup D^x$ be a database instance for schema $\mathcal{S}$, and
$\mc{Q}\!: \exists \bar{x}(P_1(\bar{x}_1) \wedge \cdots \wedge P_m(\bar{x}_m))$  a BCQ.
Assume that $\mc{Q}$ is, possibly  unexpectedly, true in  $D$. That is, for
the associated DC $\kappa(\mc{Q})\!: \forall \bar{x} \neg (P_1(\bar{x}_1) \wedge \cdots \wedge P_m(\bar{x}_m))$, it holds
$D \not \models \kappa(\mc{Q})$, i.e. $D$ violates the DC. This becomes our {\em observation}, and
we want to find causes for it, using a diagnosis-based approach, more precisely a {\em consistency-based diagnosis approach} \cite{Reiter87}.

 We consider a {\em  diagnosis problem},  $\mathcal{M}=(\nit{SD},D^n, \mc{Q})$, associated to $\mc{Q}$.
Here, $\nit{SD}$ is a FO system description (or specification)
containing the following elements: \vspace{-3mm}
\begin{itemize}
\item  [(a)] Reiter's logical reconstruction of $D$ as a FO theory \cite{Reiter82}.

\item [(b)] Sentence $\kappa(\mc{Q}){^{ext}}$, which is $\kappa(\mc{Q})$ rewritten as follows:

$\kappa(\mc{Q}){^{ext}}\!: \ \forall   \bar{x}\neg (P_1(\bar{x}_1)  \wedge \neg \nit{ab}_{P_1}(\bar{x}_1)  \wedge \cdots \wedge
P_m(\bar{x}_m) \wedge$\\ \hspace*{5.7cm} $ \neg \nit{ab}_{P_m}(\bar{x}_m) )$.


\item  [(d)] The inclusion dependencies: \
$\forall \bar{x}(\nit{ab}_P(\bar{x}) \rightarrow P(\bar{x}))$.
\end{itemize}
\vspace{-5mm}Here, predicate $\nit{ab}$ stands, as usual, for {\em abnormality}. Then, the intended meaning of $\kappa(\mc{Q}){^{ext}}$ is that {\em under normal conditions
on the tuples, the DC is satisfied}. (This predicate can be applied to endogenous tuples only.)

  Now, the last entry, $\mc{Q}$, in $\mathcal{M}$ is the {\em  observation} (or the fact that it is true), which together with the system description, {\em plus the assumption that all tuples are normal}
   (i.e. not abnormal),\footnote{That is, bringing into the theory the sentences $\forall \bar{x} \neg \nit{ab}_R(\bar{x})$.} produces  and inconsistent theory. Consequently, a {\em  diagnosis} for the diagnosis problem $\mc{M}$ is a $\Delta \subseteq D^n$, such that
\ $\nit{SD} \cup \{\nit{ab}_P(\bar{c})~|~P(\bar{c}) \in \Delta\} \cup \{\neg \nit{ab}_P(\bar{c})~|~P(\bar{c}) \in D \smallsetminus \Delta\} \cup \{\mc{Q}\}$ \ becomes consistent.

With
$\mc{D}(\mc{M},t)$ we denote the set of all subset-minimal diagnoses for $\mc{M}$ that contain  tuple $t \in D^n$.
 Similarly,  $\mc{MCD}(\mc{M},t)$ denotes the set of  diagnoses of
$\mc{M}$ that contain  tuple $t \in D^n$ and have the minimum cardinality (among those diagnoses that contain $t$).
 Clearly  $\mc{MCD}(\mc{M},t) \subseteq \mc{D}(\mc{M},t)$.

\vspace{-3mm}
\begin{proposition}
(a) Tuple $t \in D^n$ is an actual cause for $\mc{Q}$
iff $\mc{D}(\mc{M},t) \not = \emptyset$.\\
(b) For tuple $t \in D^n$, $\rho(t)=0$ iff $\mc{MCD}(\mc{M},t) = \emptyset$.
 Otherwise, $\rho(t)=\frac{1}{|s|}$, where $s \in \mc{MCD}(\mc{M},t)$.\boxtheorem
\end{proposition}
Taking advantage of results and techniques for database
repairs and  {\em consistency-based diagnosis} through hitting sets, as done in \cite{Reiter87},  it is possible to extend complexity results reported in \cite{Meliou2010a} for the causality and responsibility
problems for conjunctive queries. This is particularly the case of the problem of deciding whether a tuple is a most responsible cause for a query answer.

\section{Causes and Abduction}\label{sec:abduction}

\vspace{2mm}
Causality in databases (and everywhere) can be seen as a very fundamental concept to which many other data management notions are connected. Some of them have been mentioned above, and there are others.
We envision a broad, common framework in which these rich connections can be formulated and investigated, contributing to shed light on each of the areas involved, and most importantly,
to take advantage of each them for theoretical and computational purposes in relation to the others.

  Still from the model-based diagnosis point of view, but this time appealing to {\em abductive diagnosis} \cite{console91,EiterGL95}, it is possible to extend and formulate
  the notion of
 query-answer causality for  Datalog queries via abductive diagnosis from Datalog specifications.
So, the connection between (query-answer) causality and abduction via Datalog makes it possible to go beyond conjunctive queries (the case considered in \cite{Meliou2010a}),\footnote{In \cite{Meliou2010b} the authors apply the definition of cause as above to monotone queries.} extending causality to, e.g. recursive queries, and obtaining new results for them.
Notice that consistency-based diagnosis is usually practiced with first-order (FO) specifications, but abductive reasoning is commonly performed under logic programming approaches \cite{DeneckerK02,EiterGL97}.

A  {\em Datalog abduction problem} (DAP) \cite{EiterGL97} is of the form $\mathcal{AP}= \langle \Pi, \nit{EDB}, \nit{Hyp},  \nit{Obs}\rangle$, where: \ (a)
$\nit{EDB}$ is an input structure (a  set of ground atoms), (b) $\Pi$ is a set of Datalog rules, (c) $\nit{Hyp}$ (the hypothesis) and $\nit{Obs}$ (the observations) are finite sets of
ground atoms
with $\Pi \cup \nit{EDB} \cup \nit{Hyp} \models \nit{Obs}$.\footnote{We will assume that no predicate in a rule head of $\Pi$ appears in $\nit{EDB} \cup \nit{Hyp}$.} The elements of $\nit{Hyp}$ are the {\em abducible atoms} (or simply, abducibles), and they, in combinations, should
explain the observations. An  {\em  abductive diagnosis} (or simply, {\em a solution}) for $\mathcal{AP}$ is a minimal subset $\Delta \subseteq \nit{Hyp}$ (wrt. subset minimality), such that
$ \Pi \cup \nit{EDB} \cup \Delta \models \nit{Obs}$. We denote with $\nit{Sol}(\mathcal{AP})$ the set of abductive diagnoses for problem $\mc{AP}$.

 The {\em relevance problem} is a decision problem that naturally arises in abduction:   Given $\mathcal{AP}=\langle \Pi, \nit{EDB,Hyp},\nit{Obs}\rangle$,  and a ground
fact $h \in \nit{Hyp} $, determine whether $h$ is  {\em relevant} in $\mathcal{AP}$, i.e.  $h$ occurs in an abductive diagnosis of  $\mathcal{AP}$, denoted $h \in \nit{Rel}(\mc{AP})$.

Now, assume we are given a relational instance with $D=D^x \cup D^n$, and a Datalog
program $\Pi$ that represents a boolean, possibly recursive query. Then, $\Pi$ has a highest level zero-ary predicate $\nit{ans}$
that returns the result (or not). If $\Pi \cup D \models \nit{ans}$, we want to find actual causes and their responsibility degrees for $\nit{ans}$.
It holds that  actual causes for $\nit{ans}$\footnote{That can be defined by means of an extension of the definition of actual causes for conjunctive queries as given in
Section \ref{sec:causes}.} can be obtained from abductive diagnosis of the associated DAP $\mathcal{AP}^c:=\langle \Pi,  D^x,D^n,\{\nit{ans}\}\rangle$, where
 $\nit{ans}$ is the observation,  $\Pi \cup D^x$ is  the {\em background theory}, and $D^n$ is the set of {\em hypothesis}. More precisely, it holds: \ tuple
$t \in D^n$ is an actual cause for $\nit{ans}$ iff $t \in \nit{Rel}(\mc{AP}^c)$.

Now, in order to obtain {\em responsibilities}, consider again $\mathcal{AP}^c=\langle \Pi,D^x,D^n,\{\nit{ans}\}\rangle$, with  $\nit{Sol}(\mathcal{AP}^c)\neq \emptyset$.
$N \subseteq D^n$ is a {\em set of necessary hypothesis} if $N$ is minimal (wrt. set inclusion), such that
$\nit{Sol}(\mathcal{AP}^c_N)=\emptyset$, where $\mathcal{AP}^c_N=\langle \Pi,D^x,D^n  \smallsetminus N,\{\nit{ans}\}\rangle$. It holds: \ The responsibility of a tuple $t$ for $\nit{ans}$ is $\frac{1}{|N|}$, where $N$ is a necessary hypothesis set with minimum cardinality, such that $t \in N$.

The notion of necessary hypothesis set we just introduced extends
the notion of (single) necessary hypothesis that has been studied in
 abduction \cite{EiterGL95,EiterGL97}. The extended notion
not only captures responsibility as defined in \cite{Meliou2010a} for
Datalog queries, but it is also
interesting in the context of abduction {\em per se}. In particular, it provides
a sort of quantitative metric to rank relevant hypothesis according to their {\em degree
of necessity}, which can be captured in terms of the sizes of necessary hypothesis sets to which they belong.
 In particular, the smaller the size of a necessary
hypothesis set, the
the more necessary are its elements to be included among the relevant hypothesis that explain the
observation.  We believe
that the degree of necessity in abduction  captures the
notion of responsibility as introduced in the context of causality
\cite{cs-AI-0312038}.

\begin{example}
Consider an instance $D$ with predicates $R$ and $S$ as below, and the query $\Pi\!: \ \nit{ans} \leftarrow R(x, y), S(y)$.

\vspace{-4mm}
\begin{center} \begin{tabular}{l|c|c|} \hline
$R$  & X & Y \\\hline
 & $a_1$ & $a_4$\\
& $a_2$ & $a_1$\\
& $a_3$ & $a_3$\\
 \hhline{~--}
\end{tabular} \hspace*{1cm}\begin{tabular}{l|c|c|}\hline
$S$  & X  \\\hline
 & $a_1$ \\
& $a_2$ \\
& $a_3$ \\ \hhline{~-}
\end{tabular}
\end{center}

\vspace{-4mm}
The DAP $\mathcal{AP}^c= \langle \Pi,\emptyset, D, \{\nit{ans}\}\rangle$ (all the tuples are endogenous)
has two abductive
diagnosis:  $\Delta_1=\{S(a_1),  R(a_2, a_1) \}$ and $\Delta_2=\{S(a_3),  R(a_3, a_3)\}$. Then,
$\nit{Rel}(\mc{AP}^c)=\{ S(a_3),$ $  R(a_3, a_3),$ $S(a_1),  R(a_3, a_3)\}$.
It is easy to verify that the relevant hypothesis are the actual causes for $\nit{ans}$.

The necessary hypothesis sets of $\mc{AP}^c$ all have cardinality 2. Thus, the responsibility of each actual cause is $\frac{1}{2}$.\boxtheorem
\end{example}

In \cite{Meliou2010a}, complexity and algorithmic results for  query-answer causality apply to
conjunctive queries. Now they can be extended
by applying results and techniques for abductive reasoning, as those obtained in \cite{EiterGL97,Gottlob07}.
These extensions are a matter of ongoing research.

\section{Abduction, View Updates and\\ Repairs}\label{sec:views}

\vspace{1mm}Another direction  to explore for fruitful connections with all the above turns around  the {\em view update problem}, which is about updating a database through views.
The problem is about minimally changing the underlying database (i.e. the base relations), in such a way that the changed base instance produces the intended changes in the
view contents. Put in other terms, it is an update propagation problem, from views to base relations.
This old and important problem in databases.

Since intended changes on the views may create alternative, admissible candidates to be the updated underlying database. As a consequence, user knowledge imposed through view updates creates or reflects {\em uncertainty} about the base data.

The view update problem is related to all the problems mentioned above. We mention some connections without going into details.
First of all, the view update problem has been treated from the point of view of abductive reasoning
\cite{Kakas90,Console95}. The idea is to ``abduce" changes on base tables that explain the intended changes on the views.

The view update problem, specially in its particular form of of {\em deletion propagation}\footnote{This is the relevant case when the views are conjunctive and
capture violations of denial constraints: the violations are solved by tuple deletion, from views and base tables.}, has been recently related in \cite{Kimelfeld12a,Kimelfeld12b} to causality as introduced in
\cite{Meliou2010a}.

We also should mention that database repairs are related to the view update problem.
Actually, {\em answer set programs} (ASPs) for database repairs (c.f. \cite{monica} for a comprehensive account and references)  implicity repair the database by updating conjunctive combinations of intentional,
annotated predicates. Those logical combinations -views after all- capture violations of integrity constraints in the original database or along the (implicitly iterative) repair process
(hence the need for annotations).

Even more, in \cite{lechen}, in order to protect sensitive information, databases are explicitly and virtually ``repaired" through secrecy views that specify the
information that has to be kept secret. In order to protect
information, a user is allowed to interact only with the virtually repaired versions of the original database that result from making those views empty or
contain only null values. Repairs are specified and computed using ASP, and an explicit connection to prioritized attribute-based repairs \cite{2011Bertossi} is made \cite{lechen}.

Finally, we should note that abduction has also been explicitly applied to database repairs \cite{arieli}. The idea, again, is to ``abduce" possible repair updates that
bring the database to a consistent state.

The areas of causality for query answers, database repairs and consistent query answering, model-based diagnosis (in its consistency-based and abductive versions),\footnote{Consistency-based diagnosis and abduction-based diagnosis are not independent approaches. Connections between them have been established \cite{console91}.}  and
database updates through views will all benefit from a deeper investigations of their mutual relationships, for better understanding them, and taking advantage
of known results for some of them to obtain new results for the others.


\vspace{2mm}
\noindent {\bf Acknowledgments:} \ Research funded by NSERC Discovery, and
the NSERC Strategic Network on Business Intelligence (BIN).  Conversations on causality in databases with Alexandra Meliou during Leo Bertossi's visit to U. of Washington in 2011 are much appreciated.
He is also grateful to Dan Suciu and  Wolfgang Gatterbauer for their hospitality. Leo Bertossi also appreciates stimulating conversations with Benny Kimelfeld during his two-month
stay at LogicBlox, and the hospitality of the LogicBlox team.


{\small
\begin{thebibliography}{10}




\ignore{\bibitem{pods99}
Arenas, M., Bertossi, L. and Chomicki, J.
\newblock Consistent Query Answers in Inconsistent Databases.
\newblock {\em Proc. PODS}, 1999.}

\vspace{2mm}
\bibitem{arieli}
Arieli,~O., Denecker,~M., Van Nuffelen,~B. and Bruynooghe,~M. \newblock Coherent Integration
of Databases by Abductive Logic Programming. {\em J. Artif. Intell. Res.}, 2004, 21:245-286.

\bibitem{2011Bertossi}
Bertossi, L.
\newblock {\em Database Repairing and Consistent Query Answering}.
\newblock Morgan \& Claypool, Synthesis Lectures on Data Management, 2011.


\bibitem{lechen}
Bertossi,~L. and Li,~L. \newblock Achieving Data Privacy through Secrecy Views and Null-Based Virtual Updates. {\em IEEE TKDE}\ignore{Transaction on Knowledge and Data Engineering}, 2013, 25(5):987-1000.



\bibitem{Borgida08}
Borgida, A., Calvanese, D. and Rodriguez-Muro, M.
\newblock Explanation in DL-Lite.
\newblock {\em Proc. DL Workshop},
CEUR-WS 353, 2008.

\ignore{
\bibitem{BunemanKT01}
Buneman, P., Khanna, S. and Tan, W.~C.
\newblock Why and Where: A Characterization of Data Provenance.
\newblock {\em Proc. ICDT}, 2001, pp. 316--330.}

\ignore{
\bibitem{BunemanT07}
Buneman, P. and Tan, W.~C.
\newblock Provenance in Databases.
\newblock {\em Proc. ACM SIGMOD}, 2007,  pp. 1171--1173.
}

\bibitem{monica}
Caniupan,~M. and Bertossi,~L. \newblock The Consistency Extractor System: Answer Set Programs for Consistent Query Answering in Databases. {\em Data \& Knowledge Engineering}, 2010, 69(6):545-572.

\ignore{
\bibitem{ChapmanJ09}
 Chapman, A. and  Jagadish, H.~V.
\newblock Why Not?
\newblock {\em Proc. ACM SIGMOD}, 2009, pp.523--534.

\bibitem{Cheney09}
Cheney, J., Chiticariu, L. and  Tan, W.~C.
\newblock Provenance in Databases: Why, How, And Where.
\newblock {\em Foundations and Trends in Databases}, 2009, 1(4): 379-474.
}


\ignore{
\bibitem{Cheney11}
Cheney, J.
\newblock Is Provenance Logical?
\newblock {\em Proc. LID}, 2011.
}


\bibitem{cs-AI-0312038}
 Chockler, H.  and Halpern,  J.~Y.
\newblock Responsibility and Blame: A Structural-Model Approach.
\newblock {\em J. Artif. Intell. Res.}, 2004, 22:93-115.

\ignore{
\bibitem{Console93}
Console, L., Friedrich G., Theseider-Dupre,~D.
\newblock  On the Role of Abduction.
\newblock  {\em ACM Comput. Surv.}, 1995, 27(3): 353-355.
}

\bibitem{console91}
Console,~L., Theseider-Dupre,~D. and Torasso,~P. \newblock On the Relationship between Abduction and Deduction. {\em J. Log. Comput.}, 1991, 1(5):661-690.


\bibitem{Console95}
 Console, L., Sapino M.~L., Theseider-Dupre,~D.
\newblock   The Role of
Abduction in Database View Updating.
\newblock {\em  J. Intell. Inf. Syst.}, 1995, 4(3): 261-280.







\ignore{
\bibitem{CuiWW00}
Cui, Y., Widom, J.,  and  Wiener. J.~L.
\newblock Tracing the lineage of view data in a warehousing environment.
\newblock {\em  ACM TODS}\ignore{rans. Database Syst.}, 2000, 25(2):179-227.}




\bibitem{DeneckerK02}
 Denecker, M., and Kakas A.~C.
\newblock Abduction in Logic Programming.
\newblock In {\em Computational Logic: Logic Programming and Beyond}, 2002, LNCS 2407.


\bibitem{EiterGL95}
Eiter, T. and Gottlob, G.
\newblock The Complexity of Logic-Based Abduction.
\newblock {\em   J. ACM }, 1995, 42(1): 3-42.

\bibitem{EiterGL97}
Eiter, T.,  Gottlob, G.  and  Leone, N.
\newblock Abduction from logic programs: Semantics and complexity.
\newblock {\em  Theor. Comput. Sci.}, 1997, 189(1-2):129-177.

\ignore{\bibitem{Friedrich90}
Friedrich,~G., Gottlob.~G. and Nejdl,~W.
\newblock  Hypothesis classification, abductive diagnosis and therapy.
\newblock {\em  Proc. Internat. Workshop on Expert Systems in Engineering}, 1990, LNCS 462, pp. 69-78.
}


\bibitem{Gertz97}
 Gertz, M.
\newblock Diagnosis and Repair of Constraint Violations in Database Systems.
\newblock  PhD Thesis, Universit\"at Hannover, 1996.


\bibitem{Gottlob07}
Gottlob,~G., Pichler,~R. and Wei,~F.
\newblock Efficient Datalog Abduction through Bounded Treewidth.
\newblock   {\em  Proc. AAAI}, 2007.

\bibitem{greco14}
Greco,~S., Pijcke,~F. and Wijsen,~J. \newblock Certain Query Answering in Partially Consistent Databases. {\em PVLDB}, 2014, 7(5):353-364.

\ignore{
\bibitem{Halpern01}
Halpern, Y.~J. and Pearl,~J.
\newblock Causes and Explanations: A Structural-Model Approach.
\newblock \ {\em Proc. UAI}, 2001.
}

\bibitem{Halpern05}
Halpern, Y.~J., Pearl, J.
\newblock Causes and Explanations: A Structural-Model Approach: Part 1.
\newblock {\em British J. Philosophy of Science}, 2005,
56:843-887.

\ignore{
\bibitem{HuangCDN08}
 Huang,~J.,  Chen,~T.,  Doan,~A. and  Naughton,~J.~F.
\newblock On The Provenance of Non-Answers to Queries over Extracted Data.
\newblock {\em PVLDB}, 2008, 1(1):736--747.
}


\bibitem{Kakas90}
Kakas A.~C. and   Mancarella,~P.
\newblock  Database Updates through Abduction.
\newblock {\em Proc. VLDB}, 1990, pp. 650-661.



\ignore{
\bibitem{Karvounarakis02}
Karvounarakis, G. and Green, T.~J.
\newblock Semiring-Annotated Data: Queries and Provenance?
\newblock {\em SIGMOD Record}, 2012, 41(3):5-14.

\bibitem{Tannen10}
Karvounarakis, G.   Ives, Z.~G. and Tannen, V.
\newblock  Querying Data Provenance.
\newblock {\em Proc. ACM SIGMOD}, 2010, pp. 951--962.}



\bibitem{Kimelfeld12a}
Kimelfeld, B.
\newblock  A Dichotomy in the Complexity of Deletion
     Propagation with Functional Dependencies.
\newblock  {\em  Proc. PODS}, 2012.

\bibitem{Kimelfeld12b}
Kimelfeld,~B., Vondrak,~J. and  Williams,~R.
\newblock   Maximizing
    Conjunctive Views in Deletion Propagation.
\newblock {\em  ACM TODS}\ignore{rans. Database
    Syst.}, 2012, 7(4):24.

\bibitem{icdt07}
Lopatenko, A. and Bertossi, L. \newblock Complexity of Consistent Query Answering in Databases under Cardinality-Based and Incremental Repair Semantics. \newblock {\em Proc. ICDT}, 2007, LNCS 4353.

\bibitem{Meliou2010b}
 Meliou, A.,  Gatterbauer. W., Halpern,  J.~Y.,  Koch, C.,
 Moore K.~F. and  Suciu, D.
\newblock Causality in Databases.
\newblock {\em IEEE Data Eng. Bull}, 2010, 33(3):59-67.



\bibitem{Meliou2010a}
 Meliou, A.,  Gatterbauer, W.
 Moore, K.~F. and  Suciu, D.
\newblock The Complexity of Causality and Responsibility for Query Answers and
  Non-Answers.
\newblock {\em PVLDB}, 2010, 4(1):34-45.

\ignore{
\bibitem{Poole88}
Poole, D.
\newblock A Logical Framework for Default Reasoning.
\newblock {\em  Artificial Intelligence}, 1988, 36:27-47.
}



\bibitem{Reiter82}
Reiter, R.
\newblock Towards a Logical Reconstruction of Relational Database Theory. \newblock
 In {\em On Conceptual Modelling}, M.L. Brodie, J. Mylopoulos and J.W. Schmidt (eds.), Springer, 1984,
pp. 191-233.

\bibitem{Reiter87}
 Reiter, R.
\newblock A Theory of Diagnosis from First Principles.
\newblock {\em Artificial Intelligence}, 1987, 32(1):57-95.


\bibitem{nmr14}
 Salimi, B. and Bertossi, L.
\newblock  Causality in Databases: The Diagnosis and Repair Connections. \newblock To appear in Proc. 15th International Workshop on Non-Monotonic Reasoning (NMR 2014).
Corr Arkiv Paper cs.DB/1404.6857, 2014.

\bibitem{chomicki12}
Staworko,~S., Chomicki,~J. and Marcinkowski,~J. \newblock Prioritized Repairing and Consistent Query Answering in Relational Databases. \newblock {\em Ann. Math. Artif. Intell.}, 2012, 64(2-3):209-246.

\bibitem{ihab12}
Yakout,~M., Elmagarmid,~A., Neville,~J., Ouzzani,~M. and Ilyas,~I. \newblock Guided Data Repair. {\em PVLDB}, 2011, 4(5):279-289.



\ignore{
\bibitem{tannen}
Tannen, V. \newblock Provenance Propagation in Complex Queries.
\newblock {\em Buneman Festschrift}, 2013, LNCS 8000, pp. 483-493.
}

\end{thebibliography}

}
\end{document}